\documentclass[usenatbib]{mnras}

\usepackage{graphicx}
\usepackage{dcolumn}
\usepackage{amsmath}
\usepackage{amssymb} 
\usepackage{ctable}
\usepackage{bm}
\usepackage{epsfig}
\usepackage{epstopdf}
\usepackage{epsf,color}
\usepackage{aas_macros}
\usepackage{url}
\usepackage{widetext}

\newcommand{\cmnt}[1]{}

\newcommand{\cii}{\ion{C}{ii}}

\newcommand{\VEV}[1]{\left\langle #1 \right\rangle}

\title[Search for \cii\ Emission]{Search for C\,II Emission on Cosmological Scales at Redshift $Z\sim2.6$}

\author[A.~Pullen et al.]{Anthony R. Pullen\thanks{Email: anthony.pullen@nyu.edu},$^{1,2}$ Paolo Serra,$^{3,4}$ Tzu-Ching Chang,$^{3,4,5}$ Olivier Dor\'{e},$^{3,4}$ \newauthor and Shirley Ho$^{6,2}$  \\ 
$^1$Center for Cosmology and Particle Physics, Department of Physics, New York University, 726 Broadway, New York, NY, 10003, U.S.A.\\
$^2$McWilliams Center for Cosmology, Department of Physics, Carnegie Mellon University, 5000 Forbes Ave, Pittsburgh, PA, 15213, U.S.A.\\
$^3$Jet Propulsion Laboratory, California Institute of Technology, Pasadena, CA 91109, U.S.A.\\
$^4$California Institute of Technology, Pasadena, CA 91125, U.S.A.\\
$^5$Academia Sinica Institute of Astronomy and Astrophysics, P.O. Box 23-141, Taipei 10617, Taiwan\\
$^6$Lawrence Berkeley National Laboratory, 1 Cyclotron Road, Berkeley, CA 94720, U.S.A.}

\date{Accepted XXX. Received YYY; in original form ZZZ}

\pubyear{2017}

\begin{document}
\label{firstpage}
\pagerange{\pageref{firstpage}--\pageref{lastpage}}
\maketitle

\begin{abstract}
We present a search for \cii\ emission over cosmological scales at high-redshifts.  The \cii\ line is a prime candidate to be a tracer of star formation over large-scale structure since it is one of the brightest emission lines from galaxies.  Redshifted \cii\ emission appears in the submillimeter regime, meaning it could potentially be present in the higher frequency intensity data from the \emph{Planck} satellite used to measure the cosmic infrared background (CIB).  We search for \cii\ emission over redshifts $z=2-3.2$ in the \emph{Planck} 545 GHz intensity map by cross-correlating the 3 highest frequency \emph{Planck} maps with spectroscopic quasars and CMASS galaxies from the Sloan Digital Sky Survey III (SDSS-III), which we then use to jointly fit for \cii\ intensity, CIB parameters, and thermal Sunyaev-Zeldovich (SZ) emission.  We report a measurement of an anomalous emission $\mathrm{I_\nu}=6.6^{+5.0}_{-4.8}\times10^4$ 
$\mathrm{Jy/sr}$ at 95\% confidence, which could be explained by \cii\ emission, favoring \emph{collisional excitation} models of \cii\ emission that tend to be more optimistic than models based on \cii\ luminosity scaling relations from local measurements; however, a comparison of Bayesian information criteria reveal that this model and the CIB \& SZ only model are equally plausible. Thus, more sensitive measurements will be needed to confirm the existence of large-scale \cii\ emission at high redshifts.  Finally, we forecast that intensity maps from Planck cross-correlated with quasars from the Dark Energy Spectroscopic Instrument (DESI) would increase our sensitivity to \cii\ emission by a factor of 5, while the proposed Primordial Inflation Explorer (PIXIE) could increase the sensitivity further.
\end{abstract}
\begin{keywords}
cosmology: theory -- cosmology: observations -- large-scale structure of the universe -- ISM:molecules -- galaxies: high-redshift -- submillimeter: ISM
\end{keywords}

\section{Introduction}

Galaxy spectroscopy is one of the most vital tools in astronomy, providing information over a wide range of scales from the nature of our local neighborhood of galaxies to the evolution of the Universe.  One spectral line that has been studied over the years in this field is the fine-structure line from ionized carbon, or \cii.  Carbon, which is very abundant due to its production in stars, has an ionization energy of 11.26 eV, allowing it to be more easily ionized than hydrogen.  At gas temperatures greater than 91K, \cii\ is excited through the energy transition ${}^2P_{3/2}\to {}^2P_{1/2}$ which produces an emission line at 157.7 $\mu$m which we will refer to as \cii.  The \cii\ line is an effective tracer of star formation, in that it tends to be the brightest line in the spectra of star-forming galaxies, contributing 0.1--1\% of the far-infrared (FIR) luminosity in low-redshift galaxies.  It is also well known that the bulk of this emission tends to come from photo-dissociation regions (PDRs).  \cii\ has been detected in star-forming, local galaxies for decades, while current instruments such as the Atacama Large Microwave/Submillimeter Array (ALMA) have begun to extend detections out to high-redshift \cii\ galaxies, including one \cii\ galaxy at $z\sim 7$ \citep{2015MNRAS.452...54M} within the epoch of reionization (EoR) when early star-formation reionized the intergalactic medium (IGM).  These successes show the potential for \cii\ emission to reveal much about our local Universe.

However, the \cii\ galaxies that are detected individually at high angular resolution have been shown to be only the brightest and most massive of all \cii-emitting galaxies, characterized by a steep faint-end slope in the UV luminosity function \citep{2015ApJ...803...34B}, meaning the more representative low-mass galaxies are out of reach for these surveys.  In addition, even the most powerful upcoming survey telescopes will not produce large-scale galaxy samples past redshift $z=4$, limiting the redshifts and scales usable for cosmology.  These concerns could be rectified by producing maps of line emission at low angular resolution, a technique called \emph{intensity mapping} (IM) \citep{1990MNRAS.247..510S,1997ApJ...475..429M,1999ApJ...512..547S,2008MNRAS.383.1195W,2008PhRvL.100i1303C}.  Capturing the aggregate emission of all emitters gives us a representative picture of the properties of galaxies and star-forming regions, while also allowing us to observe directly the largest cosmological scales ($>1$Gpc) of large-scale structure (LSS), the EoR, and potentially the preceding ``dark ages.''  The IM method has been well-developed in the literature, and it has traditionally been considered in the context of mapping the 21-cm line from neutral hydrogen.  There have been a proliferation of 21-cm survey efforts, with the Canadian HI Mapping Experiment (CHIME) \citep{2014SPIE.9145E..22B}, the Hydrogen Epoch of Reionization Array (HERA) \citep{2017PASP..129d5001D}, and the Square Kilometer Array (SKA) \citep{2015aska.confE..19S} as the upcoming benchmarks for this effort.  Lately, there has been great interest in mapping other bright lines, including \cii\ \citep{2004A&A...416..447B,2012ApJ...745...49G,2015ApJ...806..209S,2015MNRAS.450.3829Y}, CO \citep{2008A&A...489..489R,2011ApJ...730L..30C,2011ApJ...741...70L,2013ApJ...768...15P,2014MNRAS.443.3506B,2015JCAP...11..028M,2016ApJ...817..169L}, Ly$\alpha$ \citep{2013ApJ...763..132S,2014ApJ...786..111P,2014ApJ...785...72G,2016MNRAS.455..725C}, and H$\alpha$ \citep{2017ApJ...835..273G,2017arXiv171109902S}.  A few surveys are also being considered to map other lines and are at various stages of development, including the CO Mapping Pathfinder \citep{2016ApJ...817..169L}, TIME (\cii) \citep{2014SPIE.9153E..1WC}, CONCERTO (\cii) \citep{serra2016}, HETDEX (Ly$\alpha$) \citep{2008ASPC..399..115H}, SPHEREx (H$\alpha$, \ion{O}{iii}, Ly$\alpha$) \citep{2014arXiv1412.4872D}, and CDIM (H$\alpha$, \ion{O}{iii}, Ly$\alpha$) \citep{2016arXiv160205178C}.  The consideration of alternatives to the 21-cm line for IM has been greatly boosted by the recent detection of CO correlations in LSS through the COPPS survey \citep{2016ApJ...830...34K}.

While IM studies are usually considered in terms of measuring auto-correlations, cross-correlating an intensity map \citep{2010JCAP...11..016V} with another tracer of LSS has advantages in that (1) instrumental noise bias is eliminated, making accessible small-scale modes which are more numerous than the large-scale modes, and (2) other lines from different redshifts in the IM map will not correlate with the other LSS tracer.  In the future, we expect to cross-correlate intensity maps from different line tracers at the same redshift \citep{2011ApJ...741...70L,2012ApJ...745...49G,2013ApJ...763..132S,2014ApJ...785...72G,2015aska.confE...4C}, \emph{e.g.} \cii\ and 21-cm, in order to track how different LSS phases are correlated.  Even now, cross-correlations have been performed between diffuse emission maps and low-redshift LSS tracers to detect the particular diffuse emission at a given redshift, including the current 21-cm detections \citep{2010Natur.466..463C,2013ApJ...763L..20M}.  Previously, \citet{2013ApJ...768...15P} performed a cross-correlation between the Wilkinson Microwave Anisotropy Probe (WMAP) temperature maps and a photometric quasar sample from the Sloan Digital Sky Survey (SDSS) II, allowing us to place limits on large-scale CO emission.  We predicted that \emph{Planck} could potentially detect \cii\ emission, which is $\sim$1000x brighter than CO emission.  Also, while contamination of \cii\ maps by thermal dust \citep{2014A&A...571A..11P} and, to a much less extent, CO emission \citep{2016ApJ...825..143L,2016ApJ...832..165C} is a concern, thermal dust and CO contamination in a cross-correlation will only increase the noise without biasing the result.

In this paper, we measure the intensity of \cii\ diffuse emission by performing an Monte Carlo Markov Chain (MCMC) analysis fitting for \cii\ and cosmic infrared background (CIB) emission jointly using cross-correlations between high-frequency intensity maps with LSS tracers.  Specifically, we measure angular cross-power spectra of overdensity maps of both spectroscopic quasars at redshift $z=2.6$ and CMASS galaxies at redshift $z=0.57$ from the SDSS-III Baryon Oscillations Spectroscopic Survey (BOSS) \citep{2011AJ....142...72E} with the \{353, 545, 857\} GHz intensity maps from the \emph{Planck} satellite \citep{2010A&A...520A...9L,2011A&A...536A...6P} to fit jointly for the \cii\ intensity and 3 CIB parameters. The spectroscopic quasars are limited to redshifts $z=2-3.2$, which comprise the redshift range of \cii\ emission within the 545 GHz band, while the redshifts of the CMASS galaxies are too low to correlate with \cii\ emission in the Planck maps.  Thus, we expect the \cii\ emission to appear only in the cross-correlation of the quasars with the 545 GHz Planck map, while the other 5 cross-correlations are used to fit the CIB parameters.

We confirm that the MCMC analysis best-fit model for the CIB and \cii\ emission constitutes a good fit to the data, with the CIB parameters in broad agreement with previous CIB analyses, and a favored value for the \cii\ intensity at $z=2.6$ of $\mathrm{I_{\cii}}=6.6^{+5.0}_{-4.8}\times10^4$ $\mathrm{Jy/sr}$ (95\% c.l.) for the \cii\ emission.  Although this is not quite a detection, it does support the possibility the high-redshift \cii\ emission is present alongside the continuum CIB emission.  Note that the favored value of $\mathrm{I_{\cii}}$ is consistent with \cii\ emission models from \citet{2012ApJ...745...49G} and \citet{2015ApJ...806..209S} that are constructed from collisional excitation models, and is in tension with lower emission models constructed from luminosity scaling relations based on local measurements.  It is possible that other extragalactic emission lines could contribute to this emission; based on local measurements, we expect $\sim$3\% of the excess cross-correlation of the Planck 545 GHz band with quasars not due to CIB to be comprised of interloping lines based on local line ratio measurements.  Using Bayesian evidence ratio, or Bayes Factor, we find that both this model and the no-\cii\ models are equally favored, and more sensitive data will be need to confirm or rule out high-redshift \cii\ emission. Finally, we forecast what the sensitivity of this measurement would be for upcoming surveys.  Replacing BOSS quasars and CMASS galaxies with luminous red galaxies (LRGs) and quasars from the upcoming Dark Energy Spectroscopic Instrument (DESI) \citep{2013arXiv1308.0847L} could increase the signal-to-noise ratio to 10.  Further replacing Planck with the proposed Primordial Inflation Explorer (PIXIE) \citep{2011JCAP...07..025K} could increase the signal-to-noise ratio to 26, in addition to allowing isolation of \cii\ emission from nearby lines due to PIXIE's high spectral resolution.


The plan of our paper is as follows: in Section \ref{S:data} we describe the \emph{Planck} and SDSS data products we use.  In Section \ref{S:analysis} we present the estimator for the intensity-LSS angular cross-power spectra, and we present our cross-power spectrum measurements and checks for systematic effects in Section \ref{S:results}.  In Section \ref{S:constraints}, we place constraints on \cii\ emission using and MCMC analysis, and in Section \ref{S:discuss} we discuss how our \cii\ constraints compare with various \cii\ models and what this implies for upcoming \cii\ surveys.  We conclude in Section \ref{S:conclude}.

\section{Data} \label{S:data}
\subsection{Planck Maps}

We use data from the \emph{Planck} satellite, which measured the intensity and polarization of the cosmic background radiation (CBR) over the entire sky.  The CBR was observed during the time period of August 2009 and August 2013 using a 74-detector array consisting of two instruments.  The Low-Frequency Instrument (LFI) \citep{2010A&A...520A...4B,2011A&A...536A...3M} implements pseudo-correlation radiometers to observe over 3 frequency channels at 30, 40 and 70 GHz.  The High-Frequency Instrument (HFI) \citep{2010A&A...520A...9L,2011A&A...536A...6P} uses bolometers and observes over 6 frequency channels at 100, 143, 217, 353, 545, and 857 GHz.

We use the 545 GHz intensity map to trace the redshifted \cii\ intensity over the sky, while additionally using the 353 and 857 GHz maps to constrain the CIB and SZ emission and clustering. The maps have beam full-widths at half-maximum (FWHMs) of the order of a few arcmin.  These maps, like all \emph{Planck} maps, use HEALPix \citep{2005ApJ...622..759G} pixelization with $N_{\rm side}=2048$.  Superimposed on these maps are one common mask constructed to remove pixels with bright Galactic emission and point sources.  Our point source veto mask is a union of all the point source veto masks for all three maps.   The \emph{Planck} Galactic emission mask with 2$^\circ$ apodization leaves 33.8\% of the sky, while our combined \emph{Planck} point source mask with 0.5$^\circ$ apodization leaves on its own 94.5\% of the sky.  Together, these two masks combined leave 33.2\% of the sky in the survey, comprising the high and low-latitude regions near the Galactic poles.  Note that we also use the 545 GHz bandpass filter to construct a radial selection function we use to predict the \cii\ emission angular power spectrum (see Fig.~\ref{F:fz}).

For the unmasked regions, we still expect emission from the cosmic microwave background (CMB) including SZ perturbations, the cosmic infrared background (CIB), and thermal dust along with the potential \cii\ emission.  We do not attempt to subtract these extra sources of emission directly.  Instead we rely on our use of cross-correlations with LSS tracers to remove the base CMB and thermal dust perturbations from our \cii\ estimator, while fitting for SZ and CIB emission simultaneously with \cii\ emission.  However, we do expect all these extra sources of emission to contribute to the parameter errors.

\subsection{BOSS Maps} \label{S:boss}

SDSS-III \citep{2011AJ....142...72E}, similar to SDSS I and II \citep{2000AJ....120.1579Y}, is constructed from galaxies and point sources detected by a 2.5 m telescope \citep{2006AJ....131.2332G} with a mounted imaging camera \citep{1998AJ....116.3040G} with five filters ($ugriz$) \citep{1996AJ....111.1748F,2002AJ....123.2121S,2010AJ....139.1628D}, that images over one-third of the sky.  Astrometric calibration \citep{2003AJ....125.1559P}, photometric reduction \citep{2001ASPC..238..269L}, and photometric calibration \citep{2008ApJ...674.1217P} is performed by automated pipelines.  Bright galaxies, luminous red galaxies (LRGs), and quasars are selected for follow-up spectroscopy \citep{2002AJ....124.1810S,2001AJ....122.2267E,2002AJ....123.2945R,2003AJ....125.2276B,2013AJ....146...32S}.  This survey took place between August 1998 and May 2013.  was the SDSS-III specifically tasked the Baryon Oscillations Spectroscopic Survey (BOSS) \citep{2013AJ....145...10D} with constructing a spatially uniform low-redshift galaxy sample and a sample of high-redshift quasars, primarily to constrain dark energy.

In our analysis we use the BOSS spectroscopic quasar sample (P\^{a}ris et al., \emph{in prep.}) from Data Release 12 (DR12) \citep{2015ApJS..219...12A} to trace LSS at $z=2.6$.  The quasars were targeted under a CORE+BONUS sample \citep{2012ApJS..199....3R}, where the CORE quasars are uniformly sampled for clustering studies and the BONUS quasars are not and mainly used to sample the Ly$\alpha$ forest.  We use only the CORE quasars in our analysis.  The target selection for the CORE quasars was implemented by applying the \emph{extreme deconvolution} (XD) technique, which determines the distribution points in parameter space, to quasars and stars in color space in order to separate their populations (XDQSO) \citep{2011ApJ...729..141B}.  The spectra of the targeted point sources are then visually analyzed to determine their spectroscopic redshifts.  A mask comprising the BOSS imaging regions is also constructed, with veto masks applied to remove areas near bright stars, centreposts of the spectroscopic plates, regions with bad photometry, particularly $u$ band data, and regions where less than 75\% of CORE targets received a BOSS spectroscopic fibre \citep{2011ApJ...728..126W}. This method has been used and is explained in more detail in previous SDSS quasar analyses \citep{2012MNRAS.424..933W,2014A&A...563A..54P,2015MNRAS.453.2779E}, and is currently being used for the Extended BOSS (eBOSS) quasar target selection \citep{2015ApJS..221...27M}.  We implement this procedure on point sources from DR12, giving us a catalogue of 178,622 quasars.  We then keep the ones in the redshift range $z=2-3.2$ in pixels with mask weight greater than 90\%, leaving us with 82,522 quasars over 8294 deg$^2$, with an overlap with the \emph{Planck} map of 6483 deg$^2$ with 75,244 quasars.

We also use the CMASS spectroscopic galaxy sample from BOSS DR12 \citep{2016arXiv160703155A,2016MNRAS.455.1553R,2015ApJS..219...12A}, which was publicly released with the final BOSS data set.  This galaxy sample does not correlate with \cii\ emission; we use it to constrain the CIB and SZ emission and clustering.  The full CMASS sample \citep{2015ApJS..219...12A} contains 862,735 galaxies over an area of 9376 deg$^2$ with a mean redshift of 0.57 and is designed to be stellar-mass-limited at $z>0.45$.  Each spectroscopic sector, or region covered by a unique set of spectroscopic tiles \citep{2011ApJS..193...29A}, was constructed to have an overall completeness, or fraction of spectroscopic targets observed, over 70\% and a redshift completeness, or fraction of observed galaxies with quality spectra, over 80\%.  We take the full CMASS sample and remove galaxies outside the redshift range $z=0.43-0.7$ and galaxies within pixels with coverage less than 90\%, leaving us with 777,202 galaxies over an area of 10,229 deg$^2$.

For both the quasars and galaxies, we construct overdensity maps $\delta_i=(n_i-\bar{n})/\bar{n}$, where $i$ is the sky pixel.  For the quasars, $n_i$ is the actual number of quasars in pixel $i$, while for galaxies $n$ is the weighted number of galaxies $n_i=\sum_{j\in {\rm pixel}\,i} w_j$ where $w_j$ is the systematic weight \citep{2014MNRAS.441...24A} of galaxy $j$.  The map is given as a HEALPix pixelization with $N_{\rm side}=1024$.  We do not attempt to weight the sky pixels by their observed areas; the HEALPix pixels are much smaller than the observed sectors that define the completeness and weighting individual pixels could introduce extra power due to possible errors in the completeness on small scales.  Finally we perform 0.5$^\circ$ apodization on both the quasar and CMASS galaxy masks.

\section{Cross-correlation analysis} \label{S:analysis}

We construct 6 angular cross-power spectra $C_\ell$, cross-correlating the 353, 545, and 857 GHz Planck maps with our BOSS quasar and CMASS galaxy samples.  Specifically, we estimate each angular cross-power spectrum in 9 band-powers of uniform width, where we use the convention introduced in \citet{2002ApJ...567....2H}
\begin{eqnarray}
\tilde{C}_b&=&\sum_{\ell}P_{b\ell}C_\ell\nonumber\\
C_\ell&=&Q_{\ell b}\tilde{C}_b\, ,
\end{eqnarray}
where
\begin{eqnarray}
P_{b\ell}=\left\{\begin{array}{ll}\frac{\ell(\ell+1)}{2\pi\Delta b}&\mbox{if $\ell\in b$};\\0&\mbox{otherwise}.\end{array}\right.\, ,
\end{eqnarray}
and
\begin{eqnarray}
Q_{\ell b}=\left\{\begin{array}{ll}\frac{2\pi}{\ell(\ell+1)}&\mbox{if $\ell\in b$};\\0&\mbox{otherwise}.\end{array}\right.\, ,
\end{eqnarray}
where $\Delta b$ is the bin size.  We consider multipoles in the range $100\leq\ell\leq1000$, which for the quasars at redshift $z=2.6$ that correlate with the \cii\ emission in the 545 GHz band, corresponds to transverse scales $k_\perp=0.023-0.23h$/Mpc.  This range allows us to avoid CMB-quasar correlations from the integrated Sachs-Wolfe effect \citep{1967ApJ...147...73S} and cosmic variance on large scales as well as nonlinear clustering on small scales.  However, nonlinear clustering should appear in this $\ell$-range for the CMASS galaxies, so we also consider in this analysis how these nonlinear scales affect our \cii\ constraints.

We estimate $\tilde{C}^{TL}_b$ between Planck map $T$ and LSS tracer map $L$ in band $b$ using a pseudo-$C_\ell$ estimator of the form \citep{2002ApJ...567....2H,2005MNRAS.358..833T}
\begin{eqnarray}\label{E:pseudocl}
\hat{C}^{TL}_b=\frac{2\pi}{b(b+1)}\sum_{b'}[\mathcal{M}^{-1}]_{bb'}^{TL}\tilde{D}_{b'}^{TL}\, ,
\end{eqnarray}
where $\tilde{D}_b^{TL}=\sum_{\ell}P_{b\ell}\hat{D}_\ell^{TL}$ is the angular cross-power spectrum of the masked maps, given by
\begin{eqnarray}
\hat{D}_\ell^{TL}=\frac{1}{2\ell+1}\sum_{m=-\ell}^\ell a_{\ell m}^T a_{\ell m}^{L*}\, ,
\end{eqnarray}
where $a_{\ell m}^T$ and $a_{\ell m}^L$ are the spherical harmonic transforms of the maps with the masked pixels set to zero. The matrix $\mathcal{M}_{bb'}^{TL}$ is given by
\begin{eqnarray}
\mathcal{M}_{bb'}^{TL}=\sum_{\ell\ell'}P_{b\ell}M_{\ell\ell'}^{TL}E_{\ell'}^TE_{\ell'}^{L}Q_{\ell'b'}\, ,
\end{eqnarray}
where $E_\ell^T=p_\ell^TB_\ell^T$ and $E_\ell^L=p_\ell^L$, $p_\ell$ and $B_\ell$ are the pixel and beam window functions, respectively, and $M_{\ell\ell'}^{TL}$ is the mode-mode coupling matrix resulting from partial sky coverage.  This matrix is given by
\begin{eqnarray}
M_{\ell\ell'}^{TL}=\frac{2\ell'+1}{4\pi}\sum_{\ell''}(2\ell''+1)W_{\ell''}^{TL}\left(\begin{array}{ccc} \ell&\ell'&\ell''\\0&0&0\end{array}\right)^2 \, ,
\end{eqnarray}
where
\begin{eqnarray}
W_\ell^{TL}=\frac{1}{2\ell+1}\sum_{m=-\ell}^\ell w_{\ell m}^T w_{\ell m}^{L*}\, ,
\end{eqnarray}
the angular cross-power spectrum of the two masks.

We analytically compute the covariance matrix using the formulas from \citet{2005MNRAS.358..833T}, modified to account for band powers as
\begin{widetext}
\begin{eqnarray}\label{E:cov}
Cov\left[\hat{C}_b^{T_1L_1},\hat{C}_{b'}^{T_2L_2}\right]&=&\frac{(2\pi)^2}{b(b+1)b'(b'+1)}\sum_{b_1b_2\ell_1\ell_2}[\mathcal{M}^{-1}]_{bb_1}^{T_1L_1}[\mathcal{M}^{-1}]_{b'b_2}^{T_2L_2}P_{b_1\ell_1}P_{b_2\ell_2}\nonumber\\&&\times\left[\frac{\mathcal{M}^{(2)}_{\ell_1\ell_2}(W^{T_1T_2,L_1L_2})C_{\ell_1}^{T_1T_2}C_{\ell_2}^{L_1L_1}\delta_{L_1L_2}}{2\ell_2+1}+\frac{\mathcal{M}^{(2)}_{\ell_1\ell_2}(W^{T_1L_2,T_2L_1})C_{\ell_1}^{T_1L_2}C_{\ell_2}^{T_2L_1}}{2\ell_2+1}\right]\, ,
\end{eqnarray}
\end{widetext}
where we use $C_\ell$s measured from the data to compute the covariance, we assume that the set of LSS tracers are from different redshifts and thus uncorrelated, and the expressions for $\mathcal{M}^{(2)}_{\ell_1\ell_2}$ are given in Eq.~27 of \citet{2005MNRAS.358..833T}.  This expression for the covariance is actually not symmetric; the asymmetry is due to the following approximation used in the derivation (see Eq.~A9 in the Appendix of \citet{2005MNRAS.358..833T})
\begin{eqnarray}
\sum_{\ell_1m_1}&C_{\ell_1}^{XY}&E_{\ell_1}^XE_{\ell_1}^YK_{\ell m\ell_1m_1}^XK_{\ell'm'\ell_1m_1}^{Y*}\nonumber\\
&&\simeq C_{\ell}^{XY}E_{\ell}^XE_{\ell}^Y\sum_{\ell_1m_1}K_{\ell m\ell_1m_1}^XK_{\ell'm'\ell_1m_1}^{Y*}\, .
\end{eqnarray}
Instead, the approximation should be agnostic with respect to $\ell$ and $\ell'$, so, following the treatment in \citet{2005MNRAS.360.1262B}, we make the replacement
\begin{eqnarray}
C^{XY}_{\ell_{1,2}}E_{\ell_{1,2}}^XE_{\ell_{1,2}}^Y\to\sqrt{C^{XY}_{\ell_1}E_{\ell_1}^XE_{\ell_1}^YC^{XY}_{\ell_2}E_{\ell_2}^XE_{\ell_2}^Y}\, ,
\end{eqnarray}
for every $C_\ell$ in Eq.~\ref{E:cov}.
Note that $C_\ell^{T_1T_2}$ includes contributions from the CIB, CMB, and thermal dust, as well as instrumental noise if $T_1=T_2$, and $C_\ell^{LL}$ includes shot noise.

\section{Results} \label{S:results}

\begin{figure*}
\begin{center}
\includegraphics[width=0.9\textwidth]{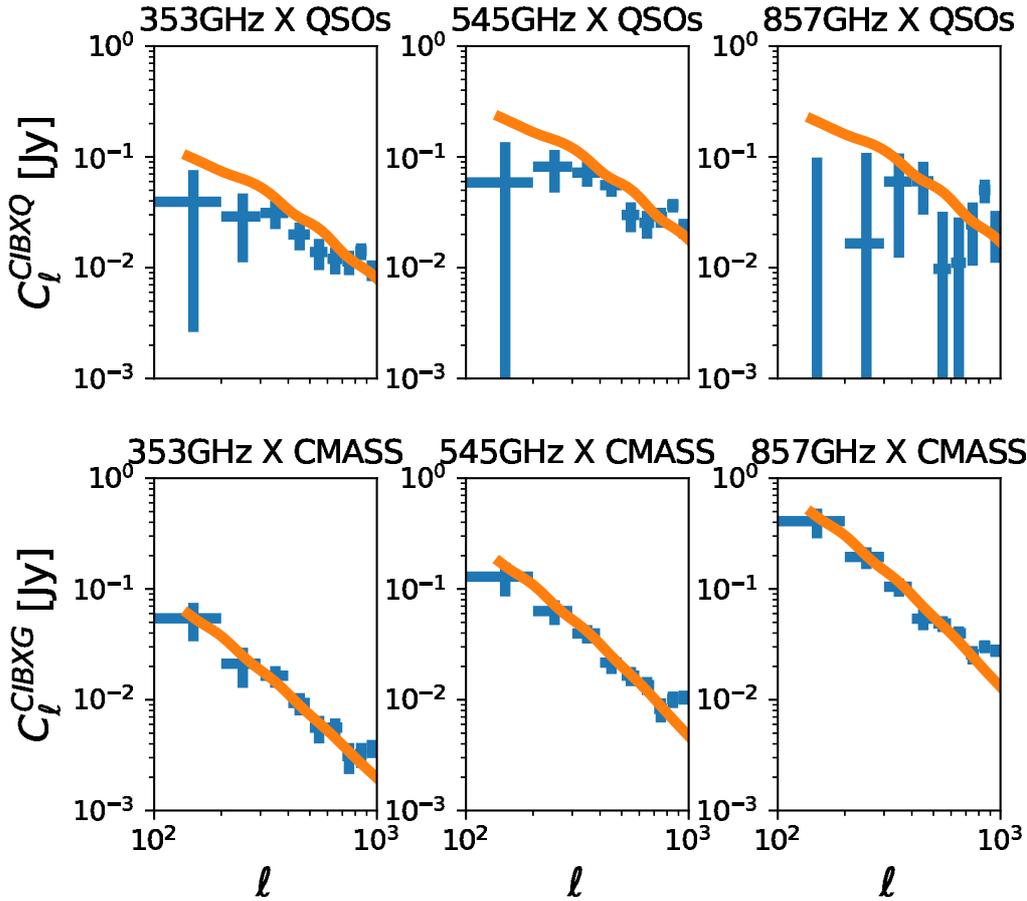}
\caption{Estimates of the six cross-power spectra from the best-fit parameters in our model, 
together with the measurements, obtained cross-correlating 
Planck CIB maps at 353, 545, 857 GHz with LRGs and QSOs.  The excesses in the Planck-LRG cross-correlations at high-$\ell$ may be due to nonlinear clustering, though we later show that these scales do not significantly affect our results.\label{F:spectra}}
\end{center}
\end{figure*}

We present in Fig.~\ref{F:spectra} our estimates of the angular cross-power spectra between the 3 Planck bands and both the quasar and CMASS galaxy samples, along with statistical errors.  The $C_\ell$s are detected with high significance, and we are able to fit them well with a sum of the CIB halo model from \citet{planckXXX2014,shang2012} and an excess due to \cii\ emission (see Sec.~\ref{S:constraints}).  Note that the $C_\ell$s for the Planck-galaxy cross-correlations agree with those presented in \citet{serra2014}.  However, these $C_\ell$s are a bit higher than the model at small scales, which may be due to nonlinear clustering.  We test the significance of these $C_\ell$s by removing the 3 highest $\ell$-bins ($\ell>700$) of the 3 Planck-galaxy $C_\ell$s from our model fits, finding that the model does not change significantly (see Sec.~\ref{S:constraints}).

\subsection{Rotation Test}
\begin{figure*}
\begin{center}
\includegraphics[width=0.9\textwidth]{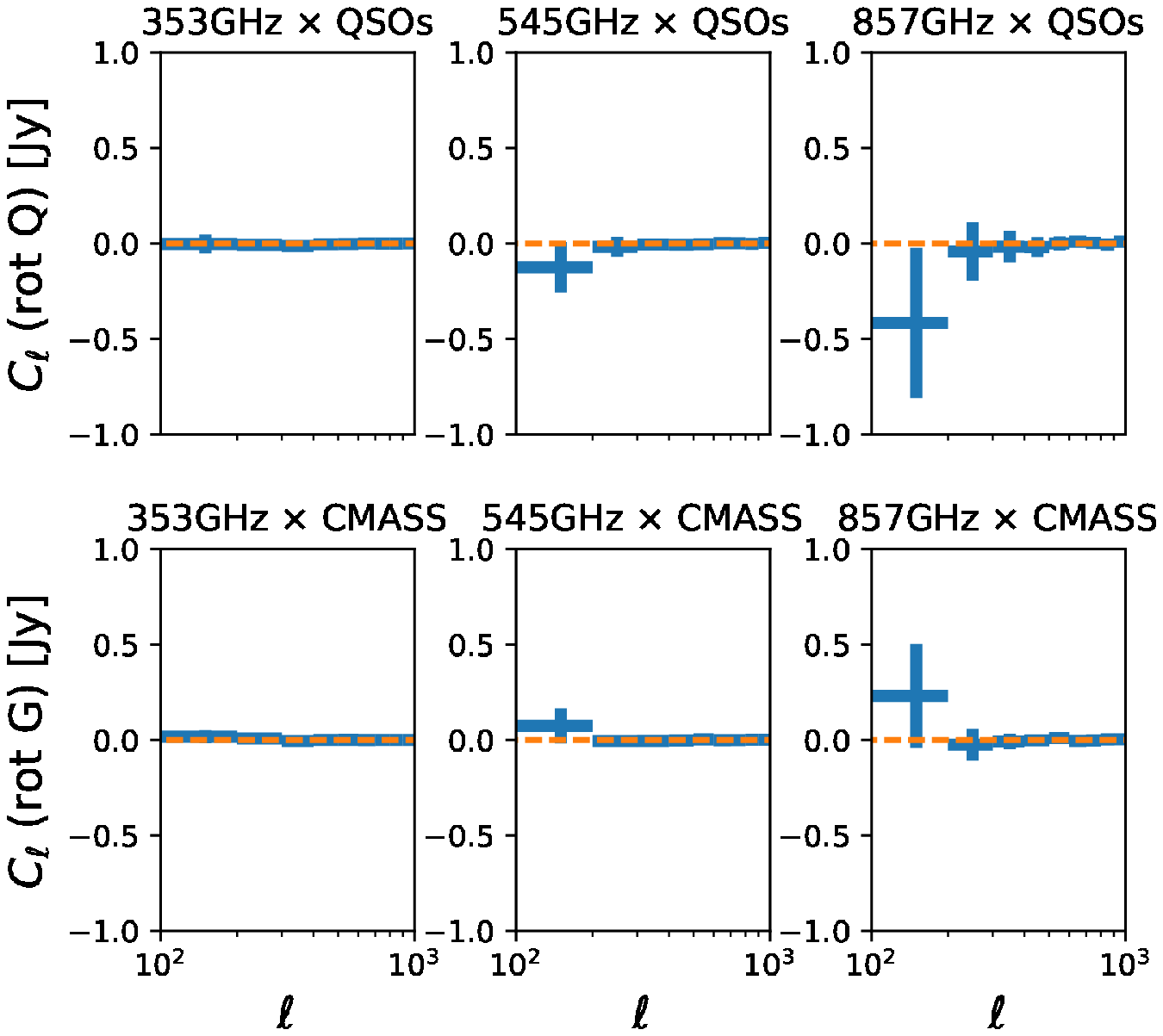}
\caption{\label{F:clrot}  The Planck-quasar (top) and Planck-CMASS (bottom) angular cross-power spectra with the quasar and CMASS maps rotated by 90$^\circ$.  The spectra appear to be consistent with a null result.}
\end{center}
\end{figure*}

We test our estimator by cross-correlating the Planck maps with LSS tracer maps rotated azimuthally $\phi\to\phi+90^\circ$.  The result, shown in Fig.~\ref{F:clrot}, has $\chi^2<2.5$ ($N_{\rm dof}=9$) for all the cross-power spectra, consistent with a null result.

\subsection{Mask Test}

\begin{figure*}
\begin{center}
\includegraphics[width=0.9\textwidth]{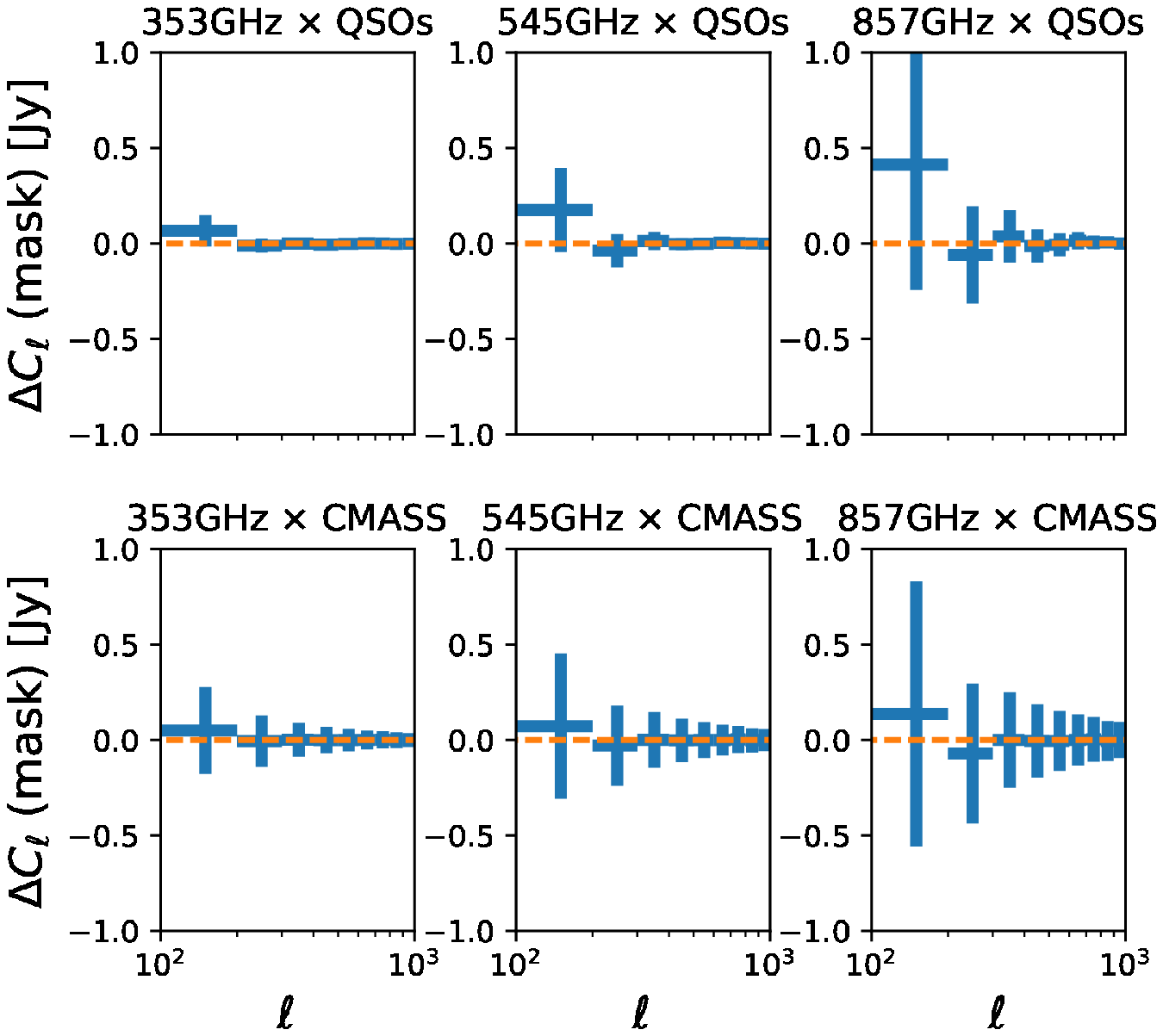}
\caption{\label{F:clmask}  The difference in the angular cross-power spectra between the 40\% Galactic dust mask, which we use for our final results, and the 20\% Galactic dust mask with 1$\sigma$ errors.  We include difference estimates for the Planck-quasar (top) and Planck-CMASS (bottom) angular cross-power spectra.  The differences for all the spectra appear to be consistent with a null result.}
\end{center}
\end{figure*}

Residual foregrounds from the Galaxy such as thermal dust and bright point sources could be correlated with systematic errors in our quasar and galaxy samples, contaminating our cross-correlations.  Since residual foregrounds are not statistically isotropic, we would expect our measurement to be dependent on the survey area if it were heavily contaminated.  In order to test this, we repeat our power spectrum measurement, replacing the 40\% Galactic mask with the 20\% Planck Galactic mask.  We then estimate the difference between our fiducial estimate (40\% Galactic mask) with the 20\% mask, which we show in Fig.~\ref{F:clmask}.  It appears that the estimate using the alternate mask is consistent with the fiducial estimate, with $\chi^2<1.5$ ($N_{\rm dof}=9$) for all the cross-power spectra, showing that our power spectrum measurement is converged with regards to the masking area.

\subsection{Jackknife Test}

\begin{figure*}
\begin{center}
\includegraphics[width=0.9\textwidth]{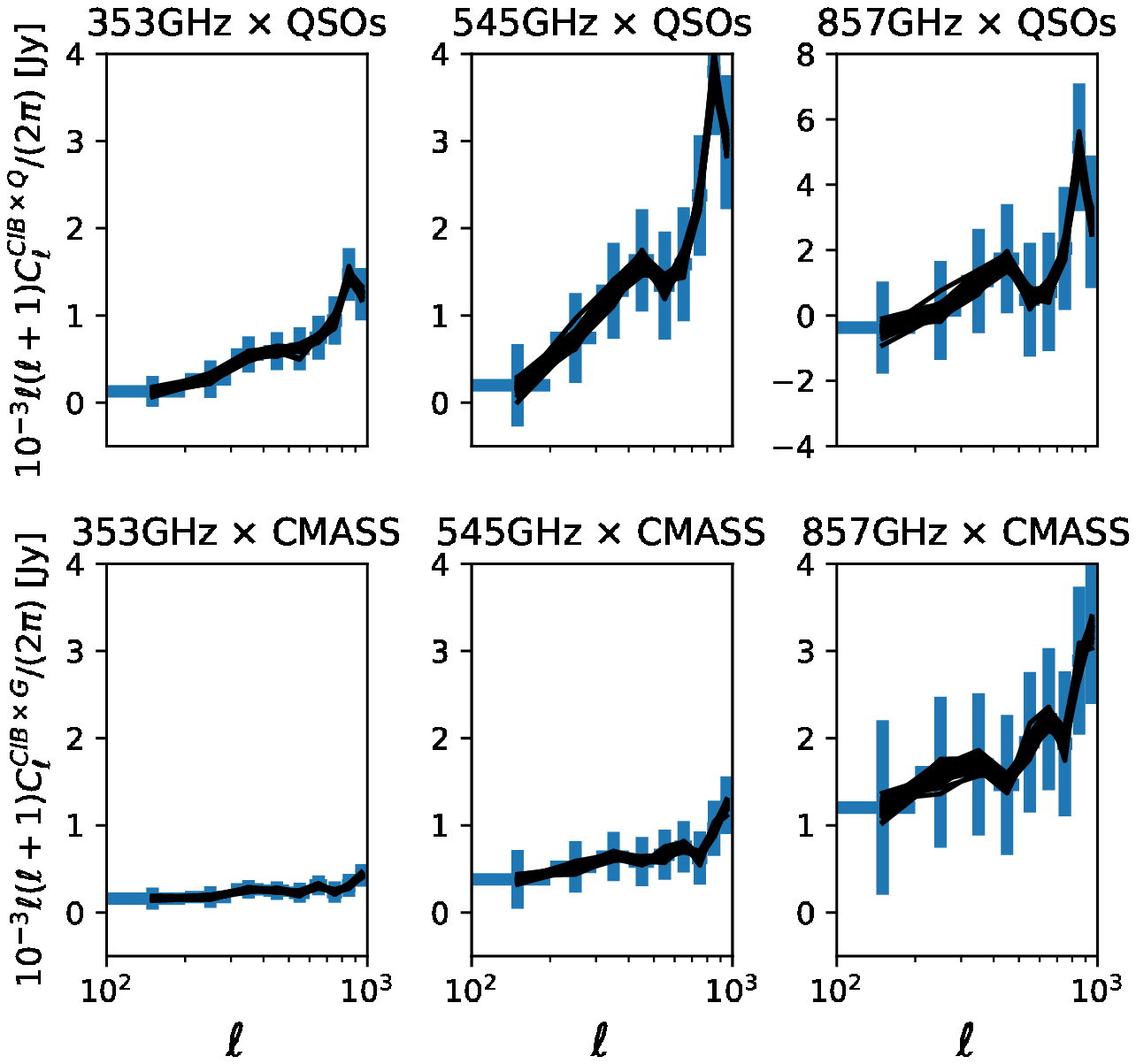}
\caption{\label{F:cljack}  The jackknife test for our Planck-quasar (top) and Planck-CMASS (bottom) angular cross-power spectrum measurements.  We construct 40 (37) jackknife regions for the quasars (galaxies), computing estimates of $C_\ell^{T-Q}$ ($C_\ell^{T-G}$), excluding each jackknife region.  The black lines are the $C_\ell$s excluding each jackknife region.  These estimates appear to be consistent with the full measurement (blue crosses), suggesting that our measurement is not dominated by foregrounds.}
\end{center}
\end{figure*}

We also test for foregrounds by performing jackknife tests, shown in Fig.~\ref{F:cljack}.  For the Planck-quasar (Planck-CMASS) power spectra, we divide both maps into 40 (37) regions, which allow us to construct multiple estimates of $C_\ell^{T-Q}$ ($C_\ell^{T-G}$), excluding each jackknife region.  This test checks that our $C_\ell$ results are not biased by foregrounds in a particular region.  For both tracers, the spread of the estimates are well within the errors, suggesting that our $C_\ell$ measurements do not vary across the sky and that foregrounds are not dominating our signal.

\section{C II Constraints} \label{S:constraints}
In order to constrain the mean amplitude of the \cii\ signal from 
CIB galaxies we fit six angular cross-power 
spectra obtained by cross-correlating 
three Planck brightness temperature, \emph{i.e.} intensity, 
maps (at $353$, $545$, $857$ GHz) with both the QSO 
overdensity map at $z\sim2.2$, and with the LRG map at $z\sim0.5$.  
Given the rest-frame wavelength of the \cii\ 
emission line at $157.7\,\mathrm{\mu}$m, and the redshift kernels 
of the Large-Scale Structure (LSS) tracers, the only observable 
containing the \cii\ line is the cross-correlation measurement 
between QSOs and the Planck temperature map at $545$ GHz. 
All the other cross-power spectra are used to constrain both 
the emission and clustering of CIB sources in the context of the halo 
model \citep{planckXXX2014,shang2012}. In this regard, we note that, while in 
principle it is possible to use measurements of the CIB auto-power spectra 
to constrain the main parameters of the model, this kind of analysis 
is complicated by the need to include many free parameters to account for the 
shot-noise power spectra of CIB anisotropies at the frequencies of interest. 
In order to keep our analysis as simple as possible, we include 
the information encapsulated in the CIB auto-power spectra simply as 
priors on the main parameters of our model of CIB galaxies.
\subsection{Cross-power spectra and the CIB model}
\begin{figure}
\begin{center}
\includegraphics[width=0.5\textwidth]{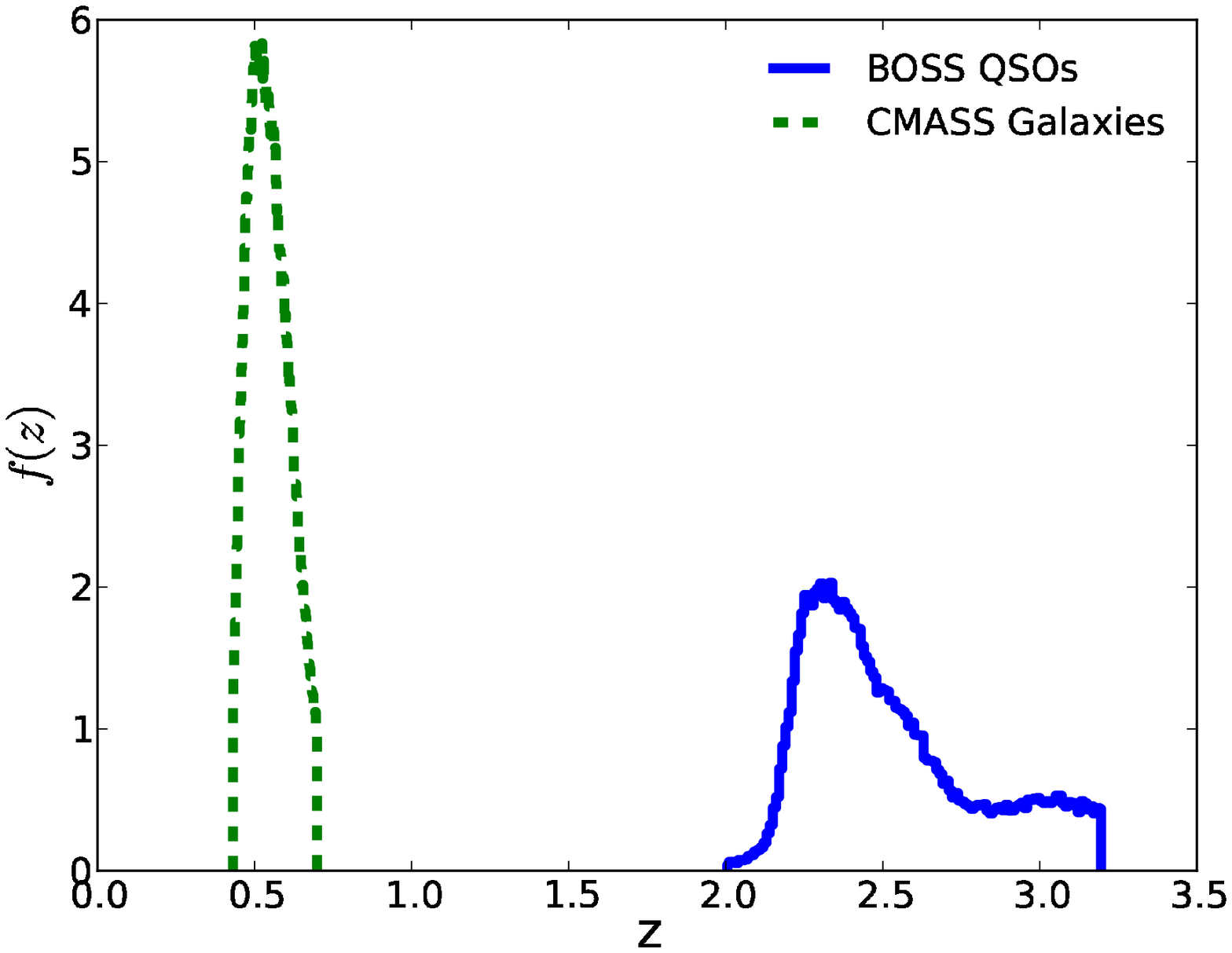}
\caption{\label{F:fz} Redshift distributions of the BOSS spectroscopic quasars and CMASS galaxies.}
\end{center}
\end{figure}

The amount of correlation between a temperature map and a generic 
LSS map (in our case a galaxy or a quasar map) 
is quantified by their cross-power spectrum, which can be expressed as:
\begin{eqnarray} \label{E:cross}
C^{\rm LSS-T}_\ell = \int\frac{dz}{\chi^2}\left(\frac{d\chi}{dz}\right)^{-1}b_{\rm LSS}b_{\rm CIB}(k,z)\nonumber\\
\frac{dN}{dz}(z)\frac{dS}{dz}(z,\nu)P_{\rm DM}(k,z)\, ,
\end{eqnarray}
where $k=\ell/\chi(z)$. The bias $\mathrm{b_{\rm LSS}}$ for LRGs is equal to 
$b_{\rm LRG}=2.1$ \citep{2017MNRAS.470.2822A}.  The BOSS quasar bias $b_{\rm QSO}$ has been measured in \citet{2012MNRAS.424..933W} to be in the range 
$b_{\rm QSO}=3.6-4.3$.  We perform our own fit for $b_{\rm QSO}$ by measuring the angular auto-power spectrum using a quadratic maximum-likelihood estimator \citep{2004PhRvD..70j3501H,2013PASP..125..705P}, finding $b_{\rm QSO}=3.5\pm0.3$.  The redshift distributions 
for both LRGs and QSOs were computed from the tracer redshift catalogs 
and are shown in Fig.~\ref{F:fz}.   

The dark matter power spectrum $P_{\rm DM}(k,z)$ is computed using \textsc{CAMB} \citep{2000ApJ...538..473L} assuming 
best-fit parameters from \cite{planck2016}. Both the 
bias $\mathrm{b_{\rm CIB}(k,z)}$ and the redshift distribution 
$\mathrm{\frac{dS}{dz}(z,\nu)}$ of CIB sources can be computed using a halo 
model for CIB anisotropies introduced in \cite{planckXXX2014,shang2012} and successfully 
applied in many subsequent analyses, 
including \cite{viero2013,planckXXX2014,serra2014,serra2016}.

In the following, we will briefly discuss the main parameters used 
in the analysis, and we refer the reader to the aforementioned 
papers for an exhaustive description of the model. The redshift distribution of CIB sources at the observed 
frequency $\nu$ can be written as:
\begin{eqnarray} \label{E:dsdzCIB}
\frac{dS_{\nu}}{dz} = \frac{c}{H(z)(1+z)}\bar{j_{\nu}}(z),
\end{eqnarray}
where the mean comoving emission coefficient $\bar{j_{\nu}}(z)$ is expressed as:
\begin{eqnarray} \label{E:jnuCIB}
\bar{j_{\nu}}(z) = \int\,dL\frac{dn}{dL}(M,z)\frac{L_{\nu(1+z)}}{4\pi}.
\end{eqnarray}
The term $\mathrm{L_{\nu(1+z)}}$ denotes the galaxy infrared luminosity 
emitted at frequency $\nu(1+z)$, and $\mathrm{dn/dL}$ is the infrared galaxy luminosity function.

The main feature of the halo model for CIB anisotropies is 
the description of the galaxy luminosity as a parametric function of 
frequency, redshift, and halo mass as:
\begin{eqnarray} \label{E:lum_gal}
L_{(1+z)\nu}(M,z)= L_0 \Phi(z) \Sigma(M) \Theta[(1+z)\nu].
\end{eqnarray}
The redshift evolution of the infrared luminosity is one of the most uncertain parameter 
in the model. We assume a power law, dependent on a single parameter $\delta$ as:
\begin{eqnarray} \label{E:phiz}
\Phi(z) = (1+z)^{\delta}.
\end{eqnarray}
The exact value of this parameter is unknown, especially at redshifts $z\ge2$, 
which is particularly relevant for our cross-correlations with quasars. 
Semianalytic models and numerical simulations predict 
different evolutions of the luminosity with redshift 
\citep{delucia2007,neistein2008,wu2016,oliver2010,bouche2010,weinmann2011}. 
For this reason, we will consider $\delta$ as a free parameter in our model.

The dependence on the dark matter halo mass is parameterized with a log-normal function as:
\begin{eqnarray} \label{E:sigmaM}
\Sigma(M) = \frac{M}{M_N} \frac{1}{(2\pi\sigma^2_{L/M})^{0.5}}\mathrm{exp}\Big[-\frac{(\mathrm{log}_{10}M - \mathrm{log}_{10}M_{\mathrm{eff}})^2}{2\sigma_{L/M}^2}\Big];
\end{eqnarray}
the term $\rm M_N$ is a normalization parameter, while $\mathrm{M_{eff}}$ 
describes the halo mass that is most efficient at hosting star formation. 
Simulations have shown that various mechanisms prevent an efficient 
star formation for halo masses much lower and much higher 
than $\mathrm{M_{eff}}$ 
\citep{benson2003,silk2003,bertone2005,croton2006,dekel2006,bethermin2012,behroozi2013}. 
We fix the value of this parameter at $\mathrm{log(M_{eff})[M_{\odot}]=12.6}$, 
in agreement with \cite{planckXXX2014, serra2016}.
The parameter $\mathrm{\sigma_{L/m}}$ accounts for the range of halo
masses mostly contributing to the infrared luminosity, and has been fixed at 
$\mathrm{\sigma_{L/m}=0.5}$ \citep{shang2012,planckXXX2014,serra2014,serra2016}.

A simple functional form
\citep[see][and reference therein]{2002PhR...369..111B} is assumed
for the galaxy Spectral Energy Distribution (SED):
\begin{eqnarray}
\Theta (\nu) \propto
\left\{\begin{array}{ccc}
\nu^{\beta}B_{\nu}\,(T_{\mathrm d})&
 \nu<\nu_0\, ;\\
\nu^{-2}&  \nu\ge \nu_0\,,
\end{array}\right.
\label{eqn:thetanu}
\end{eqnarray}
where $\mathrm{T_d}$ is the dust temperature
averaged over the redshift range considered, and
$\mathrm{\beta}$ is the emissivity of the
Planck function $\mathrm{B_{\nu}(T_d)}$. We will assume $\beta=1.5$ 
in the rest of the analysis, in agreement with \cite{planck_dust2014}.
A free parameter $\mathrm{A_{\cii}}$ is included in the fit to quantify 
the mean amplitude of the \cii\ line. At the \cii\ emission frequency 
$\nu_{\cii}=1901.03$ GHz we assume that the galaxy SED is the sum of the modified 
blackbody plus the \cii\ mean line intensity as:
\begin{eqnarray}
\Theta (\nu_{\cii}) = \Theta (\nu_{\cii})(1+A_{\cii}).
\end{eqnarray}
Note that this expression assumes that all galaxies that emit CIB will also emit the \cii\ line.  This should be a valid assumption since all galaxies should have an ionized phase and a photo-dissociation region (PDR), both of which should produce \cii\ emission.  Thus, although the intensity of the \cii\ line should vary from galaxy to galaxy, all galaxies should emit \cii\ .  In order to extract the full line information for the \cii\ emission, we could also use the Planck bandpasses as weight functions in the $C_\ell$ model.  We leave this for future work.

Finally, we must consider the additional cross-correlation 
between the Planck $353$ GHz map 
and both QSOs and LRGs, 
as due to the thermal Sunyaev-Zeldovich (tSZ) effect \citep{sunyaev1972}. 
We checked that the contamination is negligible for QSOs, but not for LRGs. We thus added 
a template describing the CIBxLRGs cross-power spectrum at 353 GHz as due to the tSZ effect, 
scaled with an amplitude $A_{\rm tSZ}$, which is a free parameter in our model. 
Based on the formalism in \citet{1999ApJ...526L...1K}, we construct the template for the SZ-LSS cross-correlation as
\begin{eqnarray}
C^{LSS-SZ}_\ell=\int\frac{dz}{\chi^2}\left(\frac{d\chi}{dz}\right)^{-1}b_{\rm LSS}\\
\frac{dN}{dz}(z)\VEV{b\frac{dy}{dz}}_{SZ}P_{\rm DM}(k,z)\, ,
\end{eqnarray}
where
\begin{eqnarray}
\VEV{b\frac{dy}{dz}}_{SZ}=\frac{dV}{dz\,d\Omega}\int dM\,n(M,z)y_\ell(M,z)b(M,z)\, ,
\end{eqnarray}
$n(M,z)$ and $b(M,z)$ are the Tinker halo mass function and halo bias \citep{2008ApJ...688..709T}, 
and $y_\ell(M,z)$ is the 2D Fourier transform of the projected Compton-y profile, given in \citet{2002MNRAS.336.1256K}.  
We then multiply the template by $1.78\times10^9$Jy/sr to convert the Compton-y parameter 
to an intensity in the Planck $353$ band based on the formula
\begin{eqnarray}
I_\nu^{SZ}=g(\nu)T_{CMB}\left(\frac{I_\nu^{SZ}}{T_{CMB}}\right)\, ,
\end{eqnarray}
where $g(\nu)T_{CMB}$ is the change in the CMB temperature due to the SZ effect, and $I_\nu/T_{CMB}$ 
is the conversion from CMB temperature to intensity, listed in Table 1 in \citet{planckXXX2014}.  This $C^{LSS-SZ}_\ell$ template is then added to the $C^{LSS-T}_\ell$ model in Eq.~\ref{E:cross} when performing the MCMC analysis.

We perform a Monte Carlo Markov Chain (MCMC) exploration of the parameter space with 
a modified version of the publicly available code 
{\tt CosmoMC} \citep{lewis2002} and using flat priors on the following set of free parameters:
\begin{eqnarray}
\Xi \equiv \{T_d, \delta, L_{\mathrm{0}}, A_{\cii}, A_{\rm tSZ}, b_{\rm QSO} \}.
\end{eqnarray}
For $b_{\rm QSO}$, we set a prior $b_{\rm QSO}=[3.2,3.8]$ based on the angular auto-power spectrum measurement discussed above.  Note that although we expect $b_{\rm QSO}$ to vary across the redshift range, we can treat it as an ``effective bias'' for our measurement since we are only searching for a redshift-independent signal, namely the \cii\ amplitude.  We were unable to perform a joint fit for $b_{\rm QSO}$ with a broad uniform prior because the CIB redshift evolution parameter $\delta$ was too degenerate with $b_{\rm QSO}$ to perform the fit without an independent measurement of $\delta$. We fit nine data points for each cross-power spectrum in the multipole range 
$\mathrm{100<l<1000}$ and, in order to obtain stronger constraints on the 
main CIB parameters, we also fit both the mean level of the CIB at 
$353$, $545$, $857$ GHz \citep{bethermin2012a} and  
a compilation of ten star formation rate density (SFRD) measurements presented 
in \cite{madau2014} and averaged over the redshift range 
$\mathrm{0<z<4}$, as in \cite{serra2016}.

With six free parameters, we are able to obtain a good fit to 
the data, with a reduced $\chi^2$ equal to $\chi^2/N_{d.o.f.}=1.3$. 
In Fig.~\ref{F:spectra} we plot the best 
fit curves obtained for the six cross-power spectra used in the analysis. 
The mean value inferred for the CIB dust temperature is 
$T_d=27.2\pm0.7$K, broadly compatible with current measurements, 
see e.g. \cite{magnelli2014}.
The redshift evolution parameter is constrained as $\mathrm{\delta}=2.3\pm0.1$, 
in agreement with constraints from previous analyses of 
CIB auto-power spectra \citep{viero2013,serra2016}, although lower than what was
found in \cite{planckXXX2014}.
The amplitude of the contamination as due to the tSZ effect,  
quantified by the parameter $A_{\rm tSZ}$, is constrained as $\mathrm{A_{\rm tSZ}}=0.75\pm0.27$.

These parameter estimates can be used to find the mean level of CIB in each of the Planck bands.  Our best-fit values for each band are 0.64 nW/m$^2$/sr (353 GHz), 2.2 nW/m$^2$/sr (545 GHz), and 5.5 nW/m$^2$/sr (857 GHz).  These values are in good agreement with results from Table 10 of \citet{planckXXX2014} and with results from \citet{bethermin2012a}, which both use auto-correlations in their measurements.  An agreement between auto-correlations and cross-correlations for the fits implies that the CIB is well-described as being fully correlated with the CMASS galaxy and quasar samples.  Since the \cii\ line should also be fully correlated with the CIB, then we can describe the \cii\ emission as being fully correlated with the LSS tracers.  If the \cii\ emission and the LSS tracers were not fully correlated, then the \cii\ constraint would be biased downward; thus, our results allow us to neglect this complication.

\begin{figure}
\begin{center}
\includegraphics[width=0.5\textwidth]{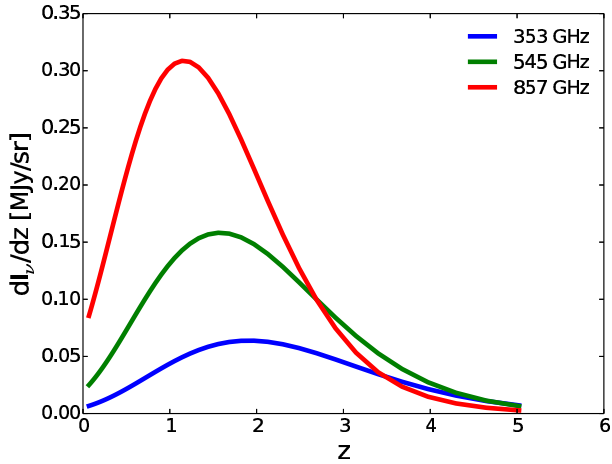}
\caption{Best-fit CIB redshift distributions for all three Planck bands.  These distributions are consistent with those from \citet{2015MNRAS.446.2696S} while also showing an expected frequency dependence due the redshifting of CIB galaxies.\label{F:didz}}
\end{center}
\end{figure}

In Fig.~\ref{F:didz} we plot the redshift distribution of the CIB based on our fit.  We find a redshift distribution consistent with that found in \citet{2015MNRAS.446.2696S} and shown in their Fig.~1.  We do see a slight discrepancy in our amplitude for the 857 GHz band distribution is a bit lower than that in \citet{2015MNRAS.446.2696S}.  Also, we claim the peak positions in our distributions show an expectant frequency dependence due to the fact that higher redshift galaxies should contribute to lower frequency bands, while the distributions in \citet{2015MNRAS.446.2696S} all have peaks in the small range of $z\simeq 1.2-1.4$.

\begin{figure}
\begin{center}
\includegraphics[width=0.5\textwidth]{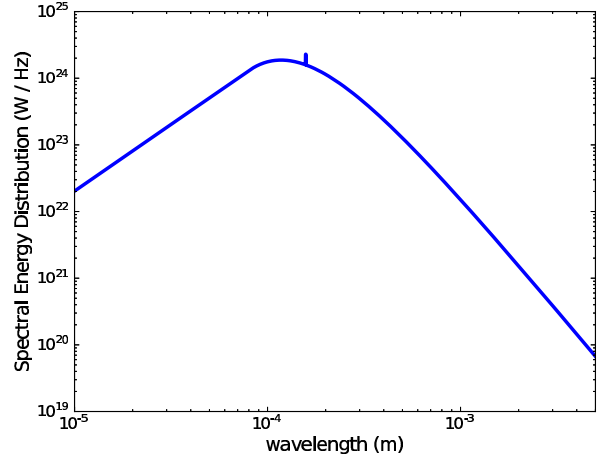}
\caption{Best-fit SED for a galaxy at $z\sim2.49$, 
where \cii\ emission at $\rm 1901$ GHz can be detected 
in cross-correlation between a QSO map and Planck's 545 GHz channel.  The SED components include a modified blackbody for the CIB and a \cii\ emission line.  The modified blackbody parameters are in agreement with previous CIB measurements \citep{planckXXX2014,bethermin2012a}.\label{sed}}
\end{center}
\end{figure}

\begin{figure*}
\begin{center}
\includegraphics[width=0.7\textwidth]{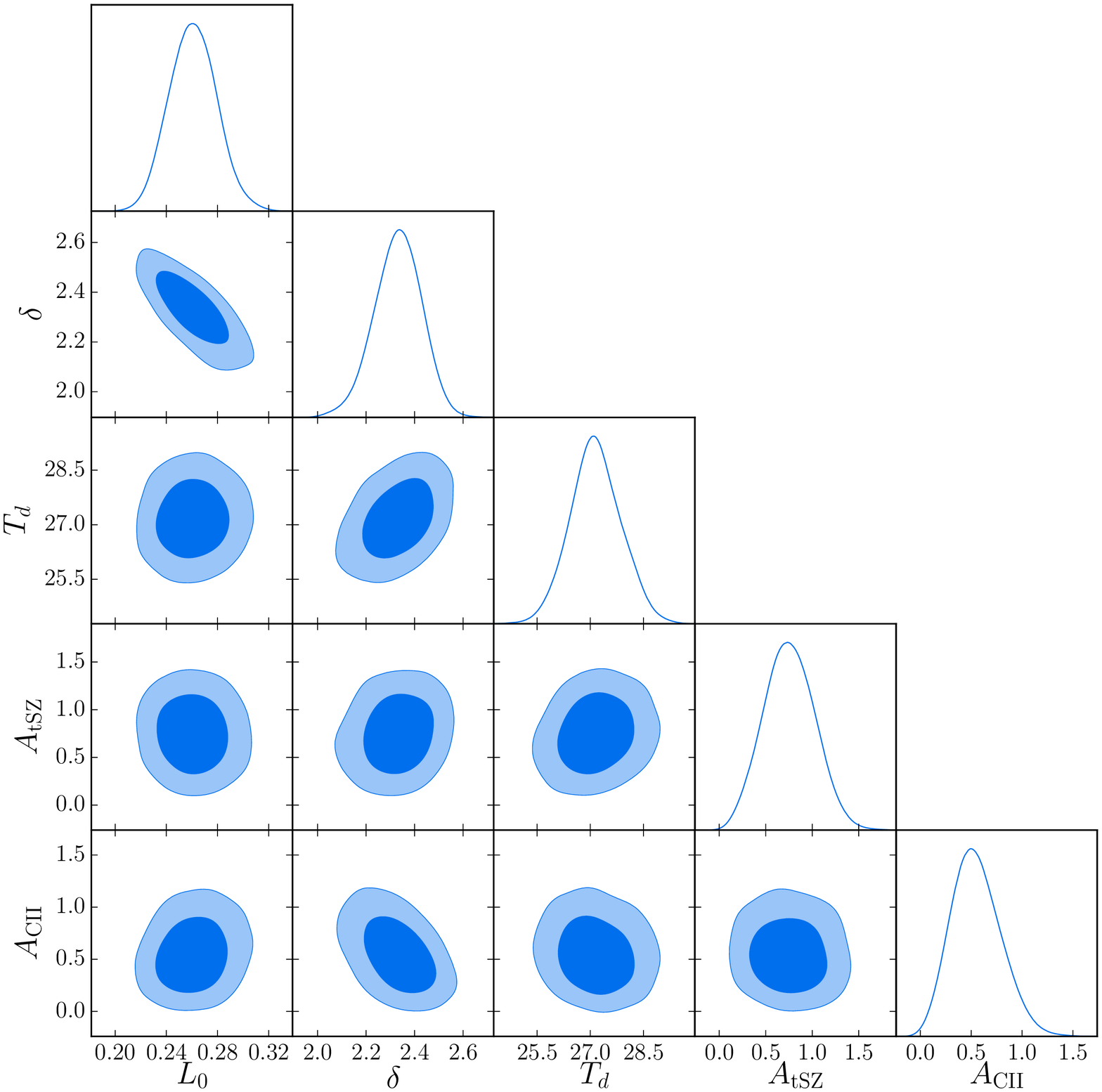}
\caption{Triangle panel showing two-dimensional confidence regions at 68$\%$ and 95 $\%$ 
for the main parameters of the model.\label{triangle_plot}}
\end{center}
\end{figure*}

Finally, the constraint on the \cii\ amplitude is set as 
$\mathrm{A_{\cii}}=0.56^{+0.42}_{-0.40}$ at $95\%$ c.l., which implies a mean 
intensity of the \cii\ line as: $\mathrm{I_{\cii}}=6.6^{+5.0}_{-4.8}\times10^4$ 
$\mathrm{Jy/sr}$ (95\% c.l.). In Fig.~\ref{sed} we present the best-fit galaxy SED with a \cii\ line, and in Fig.~\ref{triangle_plot} we show the 2-dimensional contour regions for the main parameters of the model.  Remember that we also perform this MCMC removing the 3 highest $\ell$-bins ($\ell>700$) of the 3 Planck-galaxy $C_\ell$s to test for bias due to nonlinear clustering.  This result is unchanged relative to the fit using all the $\ell$-bins.

\subsection{Contaminating Spectral Lines}

\cii\ is not the only emission line that could be present in the Planck maps.  All 3 Planck bands we use in our fit should be contaminated by lines other than \cii\, and many of them should appear at the right redshifts to correlate with either the BOSS quasars or the CMASS galaxies, biasing our CIB and \cii\ measurements.  The major ones include \ion{O}{i} (145 $\mu$m) and \ion{O}{iii} (88 $\mu$m) for the correlations with BOSS quasars and \ion{N}{ii} (205 $\mu$m) for the correlations with the CMASS galaxies, although much of the signal also comes from fainter lines.

To estimate how biased are our CIB and \cii\ estimates, we compute the angular cross-power spectra with the all the lines from Table 1 of \citet{2011JCAP...08..010V} included to see how much the amplitudes of the spectra are affected.  We use the best-fit value for $\mathrm{A_{\cii}}$ to compute the \cii\ contribution, and then scale the contributions from the other lines based on their listed luminosity-to-star-formation ratios $L/SFR$ in \citet{2011JCAP...08..010V}.  Specifically, the SED $\Delta\Theta$ added to the clean SED $\Theta$ at the rest frequency of line $X$ ($\nu_X$) is
\begin{eqnarray}
\Delta\Theta(\nu_X)=\mathrm{A_{\cii}}\Theta(\nu_{\cii})\left(\frac{(L/SFR)_X}{(L/SFR)_\cii}\right)\, .
\end{eqnarray}
This is used to then calculate the total $C_\ell^{T-LSS}$ in Eq.~\ref{E:cross} including all the lines.  We then subtract $C_\ell^{T-LSS}$ without interlopers to get the interloper contribution, $\Delta C_\ell^{T-LSS}$.  Note that we do not consider distortions due to projection effects \citep{2016ApJ...825..143L} because $C_\ell$ is already projected along the line of sight, such that even upon a shift in redshift,  fluctuations per pixel are conserved.

\begin{figure*}
\begin{center}
\includegraphics[width=0.7\textwidth]{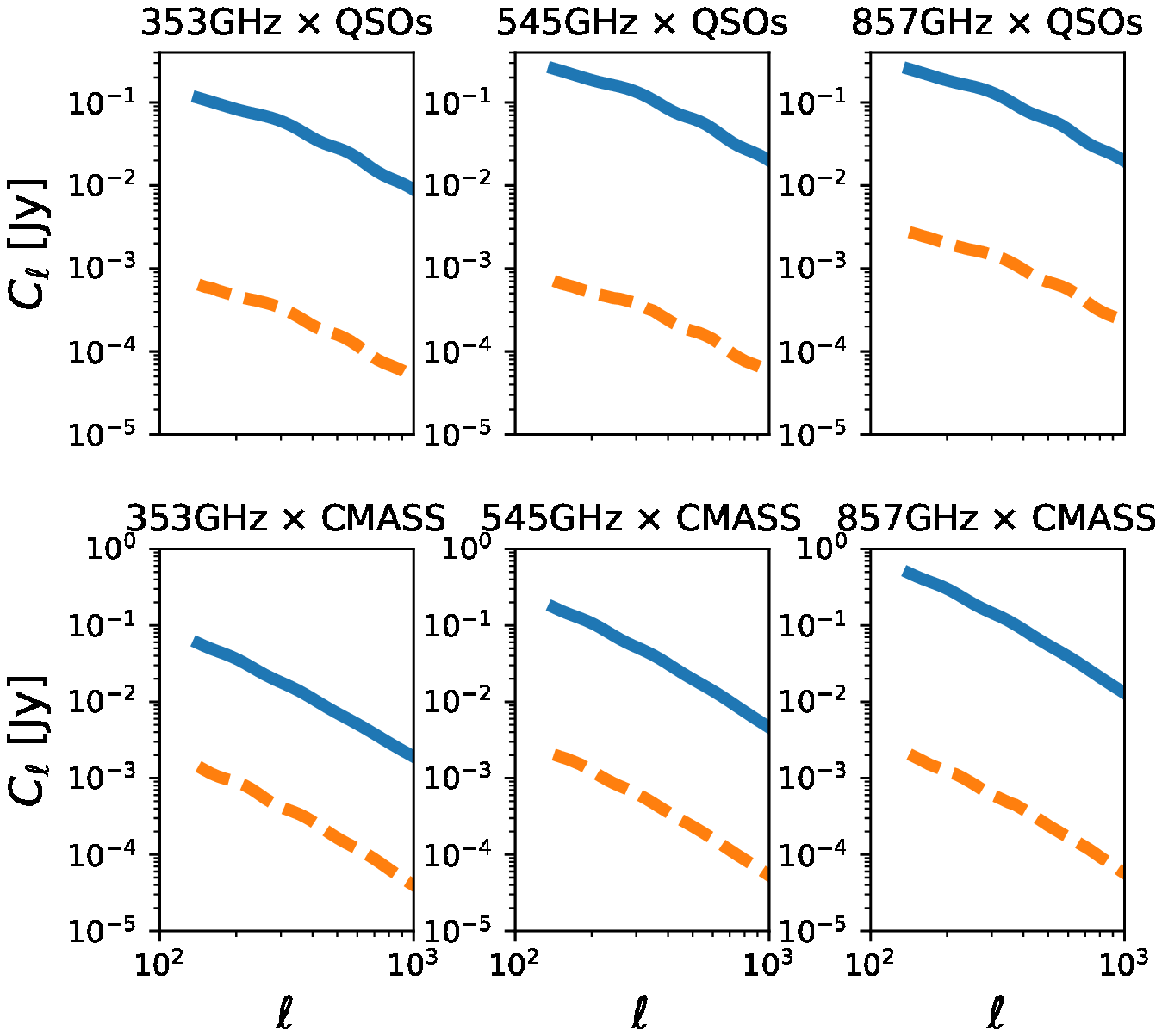}
\caption{The contribution of interlopers to the measured angular cross-power spectra.  We plot $C_\ell^{T-LSS}$ (solid) and $\Delta C_\ell^{T-LSS}$ (dashed) for all 6 cross-correlations.  We see that the interlopers contribute negligibly to the power spectra compared to our measurement errors.\label{F:clint}}
\end{center}
\end{figure*}

\begin{figure}
\begin{center}
\includegraphics[width=0.5\textwidth]{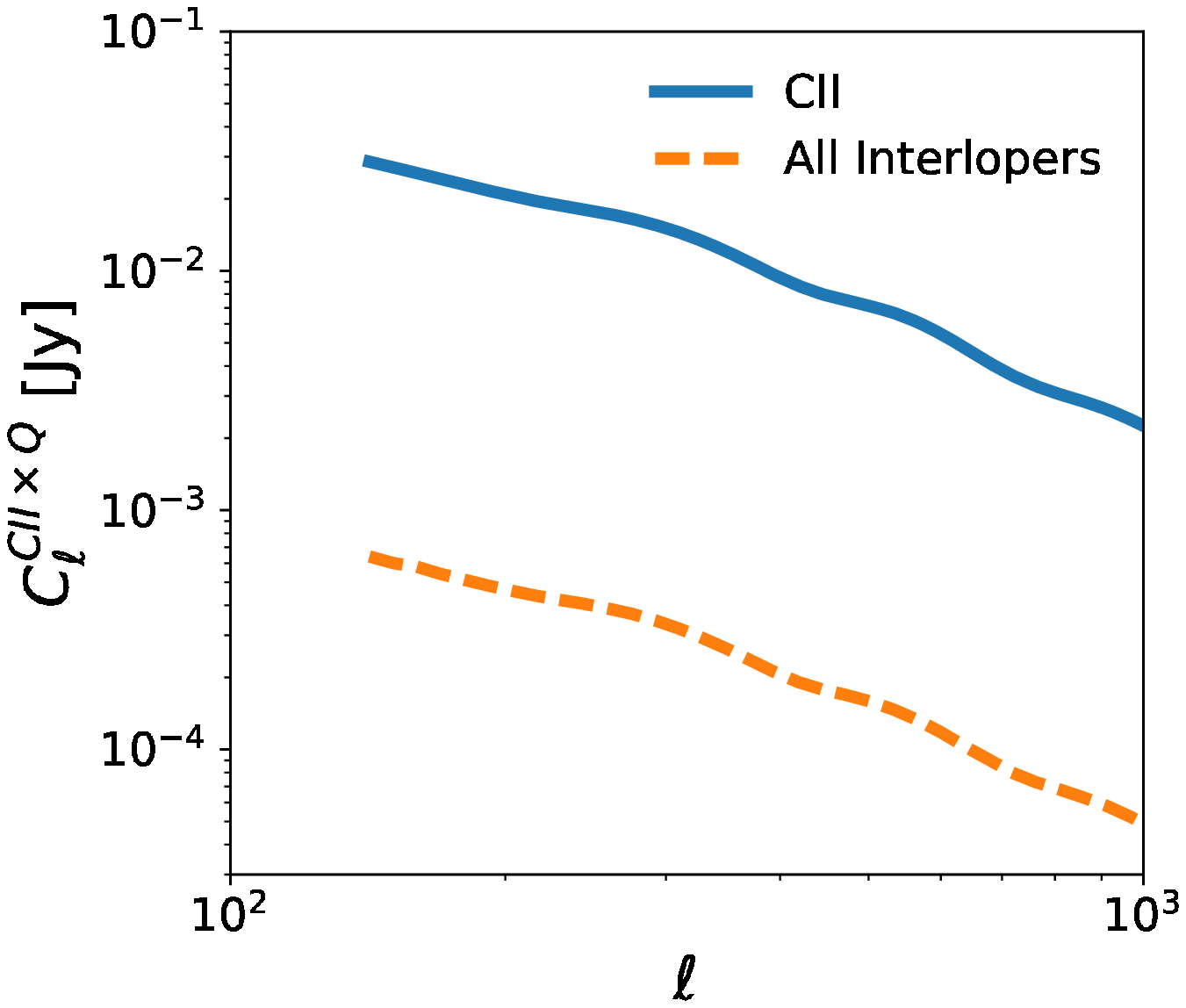}
\caption{The contribution of interlopers to the constructed angular \cii\-QSO cross-power spectrum.  We plot $C_\ell^{\cii\-Q}$ (solid) and $\Delta C_\ell^{\cii\-Q}$ (dashed) for all 6 cross-correlations.  We see that the interlopers contribute negligibly to the power spectra compared to our measurement errors.\label{F:clc2qint}}
\end{center}
\end{figure}

In Fig.~\ref{F:clint} we plot all 6 cross-correlations, including both $C_\ell^{T-LSS}$ (no interlopers) and $\Delta C_\ell^{T-LSS}$.  We plot $C_\ell^{CII-Q}$ (no interlopers) and $\Delta C_\ell^{CII-Q}$ in Fig.~\ref{F:clc2qint}.  We can see that the interloper contribution to the total $C_\ell^{T-LSS}$ is indeed subdominant.  The results for all 6 cross-correlations, plus the \cii-quasar cross-correlation, are listed in Table \ref{T:bias}.  We find that most of the correlations change by less than 2\% when including the interlopers, while $C_\ell^{353-G}$ increases by 2.3\% and $C_\ell^{\cii-Q}$ increases by 2.5\%.  Thus, we expect that the CIB and \cii\ measurements are biased by less than 3\%, which is significantly less than our measurement errors.  We do caution, however, that the line ratios used were measured using low-redshift galaxies and may not be fully accurate.  It should also be noted that the luminosity-to-star-formation ratios in \citet{2011JCAP...08..010V} were calculated using different different sets of galaxies, so there also could be mis-calibrations.  We do not expect the line intensities to possibly be of high significance for our measurements, so we do not consider it further.

\begin{table}
\begin{center}
\caption{\label{T:bias} Bias to Planck-LSS cross-power amplitudes due to interloping spectral lines, as well as the bias to the \cii\-quasar cross-correlation.  The values were derived using luminosity-to-star-formation ratios from Table 1 of \citet{2011JCAP...08..010V}.  The luminosity-to-star-formation ratios mostly come from calculations by \citet{2008A&A...489..489R} using low-redshift galaxies \citep{2001ApJ...561..766M} and measurements of the galaxy M82 \citep{2010A&A...518L..37P}.  These biases are insignificant relative to our errors.}
\begin{tabular}{ccc}
\hline
$C_\ell$&Interlopers&$\Delta C_\ell/C_\ell$[\%]\\
\hline
353-QSO&$^{12}$CO(10-9),$^{12}$CO(11-10),&0.55\\
&$^{12}$CO(12-11)&\\
545-QSO&OI&0.28\\
857-QSO&OIII&1.1\\
353-CMASS&$^{12}$CO(5-4),$^{13}$CO(5-4),HCN(6-5)&2.3\\
545-CMASS&$^{12}$CO(7-6),$^{12}$CO(8-7),CI,&1.2\\
&$^{13}$CO(7-6),$^{13}$CO(8-7)&\\
857-CMASS&$^{12}$CO(11-10),$^{12}$CO(12-11),NII&0.44\\
\cii-QSO&all interlopers&2.5\\
\hline
\end{tabular}\end{center}
\end{table}

\section{Discussion} \label{S:discuss}
Our analysis implies a non-zero amplitude of the mean \cii\ emission line at more 
than 95$\%$ confidence level. Taken at face value, this would be the first 
measurement of the \cii\ line from Planck's temperature maps. 
Because the model used to fit the data is quite uncertain (especially in the 
redshift range relevant for the cross-correlation between temperature maps and quasars) 
it is interesting to ask whether such a detection is real or due to our ignorance of the
exact values of some key parameters. In the context of Bayesian model selection, it is possible to assess the need to include the \cii\ amplitude in the fit by computing the Bayesian evidence ratio, or Bayes Factor $B$ \citep{2009arXiv0906.0664H}, using the best-fit likelihood values obtained from the MCMC fits to the data with and without a free \cii\ amplitude.  We compute $B$ as the evidence for the non-\cii\ model vs.~ the \cii\ model.  Using the Laplace approximation, the expression for the Bayes Factor in this case is given by
\begin{eqnarray}
B=\sqrt{\frac{{\rm det}\,\mathbf{C}'}{2\pi\,{\rm det}\,\mathbf{C}}}\exp\left(-\frac{1}{2}\delta\theta_\alpha[\mathbf{C}^{-1}]_{\alpha\beta}\delta\theta_\beta\right)\Delta\mathrm{A_{\cii}}\, ,
\end{eqnarray}
where $\theta_\alpha$ label the six parameters in the MCMC fit including $\mathrm{A_{\cii}}$, $\delta\theta_\alpha$ are the differences between the best-fit parameter values among the two MCMC fits, $\Delta\mathrm{A_{\cii}}=10$ is the prior on $\mathrm{A_{\cii}}$, and $\mathbf{C}$ and $\mathbf{C}'$ are the covariance matrices for the parameters for the MCMC fit with and without a free $\mathrm{A_{\cii}}$, respectively.  Note that in the fit without \cii\ emission, $\mathrm{A_{\cii}}=0$ such that $\delta\theta(\mathrm{A_{\cii}})=\mathrm{A_{\cii}}$.  We evaluate a value for the Bayes Factor of $B=1.48$; this slightly favors the model without \cii\ emission, but the rule-of-thumb is that a result $B\lesssim3$ is inconclusive.   We thus see that the data is not discriminative enough to allow an assessment of the need for the \cii\ parameters in the fit.  More sensitive measurements will be needed in the future to discriminate between the \cii\ and no-\cii\ models.

Although we cannot assert a detection of \cii\ emission, we can place constraints on \cii\ models under the assumption that the \cii\ emission does exist.  We consider several models in relation to our constraints.  The first model (Gong12) uses the values from \citet{2012ApJ...745...49G}, which predicts the \cii\ intensity from collisional excitation models as a function of the kinetic temperature and number densities of electrons, $T_k^e$ and $n_e$.  We also include a modified version of Gong12 where (1) we change the cosmological parameters to match those from joint \emph{Planck}-BOSS constraints \citep{2014A&A...571A..16P,2016arXiv160703155A} and (2) we replace the quantity $f_{\rm gas}^{\rm cr}\overline{n}_b(z)$ in the Gong12 model, where $f_{\rm gas}^{\rm cr}$ is the fraction of gas in collapsed halos and $\overline{n}_b(z)$ is the average number density of baryons, with
\begin{eqnarray}
f_{\rm gas}^{\rm cr}\overline{n}_b(z) = \frac{\Omega_b}{\Omega_m}\frac{1-Y_{\rm He}}{m_p}\rho_{\rm halo}(z)\, ,
\end{eqnarray}
where $\Omega_b$ and $\Omega_m$ are the relative baryon and matter cosmological densities, respectively, $Y_{\rm He}$ is the helium mass fraction, and $m_p$ is the proton mass.  We then consider two models from \citet{2015ApJ...806..209S}.  One model (S15M) updates the Gong12 model with recent metallicity simulations \citep{delucia2007,2011MNRAS.413..101G}.  The other model (Silva15L) uses various low-redshift luminosity measurements to construct a $L_\cii-\psi$ relation, where $\psi$ is the star formation rate, with the star formation rate constructed using the previously mentioned simulations \citep{delucia2007,2011MNRAS.413..101G}.  For models Gong12 and Silva15M the range shown is somewhat based on the range of $T_k^e$ and $n_e$ values considered in \citet{2012ApJ...745...49G}.  While the lower range corresponds to $T_k^e=10^2$K and $n_e=1$cm$^{-3}$, the upper range corresponds to the highest possible value allowed in the model, where $T_k^e=n_e\to\infty$ sets an infinite spin temperature for the \cii\ transition.  Next, we consider a \cii\ emission model by \citet{2015MNRAS.450.3829Y}, which we call Yue15.  This model constructs a $L_\cii(\psi,Z)$ fitting formula, where $Z$ is the galaxy metallicity.  The metallicity model used is dependent on the stellar mass within the halo, which we attain for the halos using results from \citet{behroozi2013}.  We also consider a model (Serra16) given in \citet{serra2016}.  We can separate these models into 2 sets: the Gong12 and Silva15M models are \emph{collisional excitation models} where the \cii\ intensity is produced by collisional excitations of \cii\ ions and electrons, and the other models are \emph{scaling relations} where the \cii\ intensity is modeled based on measured luminosity functions or SFRD measurements at low redshifts.  Note that the range of the models given in \citet{2017MNRAS.464.1948F} comprises the predictions of the scaling relation models.  The collisional excitation models tend make higher predictions than the scaling relation models.  Of course, changes in other model and clustering parameters can change the predictions. 

\begin{figure}
\begin{center}
\includegraphics[width=0.5\textwidth]{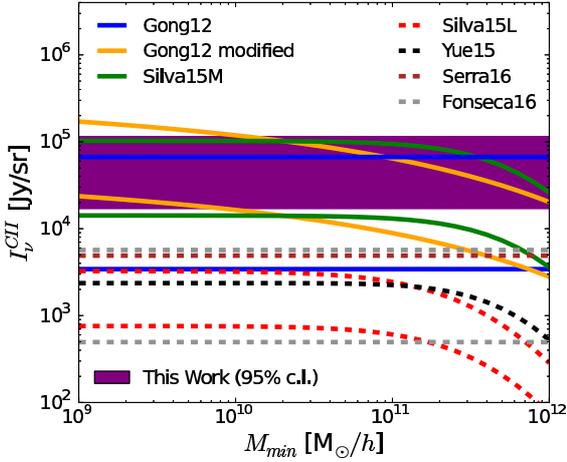}
\caption{\label{F:acii} Measurement of the quantity \cii\ intensity with 95\% confidence limits.  We also show the range of predictions for several \cii\ intensity models, including collisional excitation models (solid lines) and scaling relations (dashed lines), as functions of minimum halo mass $M_{\rm min}$ (see text for details).  Our measurement favors the collisional excitation models which appear at the high end of the range of models, although no models are ruled out by 3$\sigma$.}
\end{center}
\end{figure}

In Fig.~\ref{F:acii}, we show our \cii\ intensity constraints along with predictions based on these 7 models for \cii\ emission.  We find that our constraints favor the collisional excitation models, although none of the models shown are ruled out, in that our $A_{\cii}$ measurement is not 3$\sigma$ away from zero.  In addition, more measurements need to be performed to rule out any foreground contamination which could bias our results.  Note that the Gong12 model assumes that the ground state fraction of \cii\ ions is 1/3 for all \cii\ spin temperatures, which is actually not valid at the low intensity end when the spin temperature is much less than the CMB temperature.  This is why the spread in the Gong12 model is much larger than that of the modified Gong12 and Silva15M models where we use the spin-temperature-dependent ground fraction.  These two models are well within our constraints for reasonable values of $T_k^e$ and $n_e$, although our constraints favor $M_{\rm min}<10^{11}M_\odot/h$ for \cii\ emission.  These models are subject to improvement as we get more higher-redshift measurements of \cii\ luminosities.
     
\subsection{Forecasts for upcoming surveys}

Looking forward, we consider how sensitive upcoming surveys could be to \cii\ emission using cross-correlations.  First, we replace BOSS spectroscopic quasars with quasars from the upcoming Dark Energy Spectroscopic Instrument (DESI) \citep{2013arXiv1308.0847L}.  In the redshift range $z=2-3.2$, DESI will observe 6x more quasars than the BOSS sample we used in our analysis.  We also assume we will use the DESI LRG sample to help constrain the CIB emission.  We assume that dust and CMB emission, as in our measurement, will not be subtracted from the Planck maps.

To forecast the \cii\ sensitivity, we perform a Fisher calculation of the errors over 6 cross-correlations between the 3 Planck bands and the DESI LRGs and quasars, assuming the best-fit values from our measurement.  We first confirm that we could reproduce the sensitivity of our current measurement with a Fisher analysis, and we find that our Fisher errors are close to those found from the MCMC.  The Fisher error for the \cii\ intensity is about 1.3x the MCMC error, which is reasonable.  We predict the signal-to-noise ratio (SNR) of the \cii\ intensity for the Planck/DESI configuration to be 10, or \~5x greater than that from our measurement.  This type of measurement would be able to confirm or rule out our \cii\ measurement.

In addition, we consider detecting \cii\ emission in the proposed Primordial Inflation Explorer (PIXIE) \citep{2011JCAP...07..025K,2017ApJ...838...82S} by cross-correlating its intensity maps with maps of luminous red galaxies (LRGs) and quasars from the upcoming Dark Energy Spectroscopic Instrument (DESI) \citep{2013arXiv1308.0847L}.  For PIXIE, we assume the specifications given in \citet{2015PhRvL.115z1301H}, and we also assume that the DESI footprint is totally contained within the PIXIE footprint.  The PIXIE spectrometer has much higher spectral resolution thank Planck, with bandwidths of 15 GHz over the range of 30-1230 GHz.  Because of this, we assume that dust and the CMB will be able to be subtracted directly from the maps, allowing us to remove them from the statistical noise.  In order to do a straightforward comparison with our measurement, we group the relevant channels into the 353, 545, and 857 GHz bands from Planck and increase the band sensitivities by $\sqrt{N_{\rm channel}}$, while also considering areal number densities of LRGs and quasars over the same redshift ranges as we used for the CMASS galaxies and BOSS quasars in our measurement.  Also, PIXIE has a angular beam size of 1.6$^\circ$, much larger than Planck, so we only consider modes $100<\ell<512$.

We also perform a Fisher calculation of the errors over 6 cross-correlations, this time between the 3 (simulated) bands using PIXIE channels and the DESI LRGs and quasars, assuming the best-fit values from our measurement.  Note that we set up our forecasts to comprise the same redshift range as our measurement, though we consider a subset of the comoving scales.  We find that the sensitivity of the PIXIE/DESI configuration is 26, or  \~13x greater than that from our measurement.  In addition, the high spectral resolution of PIXIE should make it better equipped to remove interlopers by cross-correlating the LRG and quasar samples with individual PIXIE channels.  Assuming an \cii\ intensity the same as our measurement, this should be strong enough to make relevant constraints on the kinetic temperature and number density of electrons in the photo-dissociation regions powering the \cii\ emission.  Also, a \cii\ intensity of the measured magnitude extrapolated to $z\simeq6$ based on the Gong12 model could have an intensity approximately equal to those in the forecasts for TIME-Pilot \citep{2014SPIE.9153E..1WC}, which predicted a SNR of $\simeq7$.  Although the SNR may vary from this value due to uncertainties in the redshift evolution over this range, we should still learn more about the physical processes behind \cii\ emission through these measurements.

\section{Conclusions} \label{S:conclude}

We place the first constraints on \cii\ emission at large scales and redshifts $z=2-3$ using cross-power spectra between high-frequency Planck intensity maps and both spectroscopic quasars and CMASS galaxies from SDSS-III. We find $\mathrm{I_{\cii}}=6.6^{+5.0}_{-4.8}\times10^4$ 
$\mathrm{Jy/sr}$ (95\% c.l.), which favors collisional excitation models, such as the Gong12 \citep{2012ApJ...745...49G} and Silva15M \citep{2015ApJ...806..209S} models over models from luminosity scaling relations, though neither are ruled out.  In addition lower values for the minimum \cii\ -emitting halos are also favored, specifically $M_{\rm min}<10^{11}M_\odot/h$.  We found that the contribution from interloping lines are small compared to measurement errors.  The no-\cii\ model is equally plausible based on the data, and if confirmed through more sensitive measurements, this emission could also be (partially) due to other lines, or some unknown systematic.  More sensitive measurements are needed to confirm this extragalactic signal and rule out foreground contamination, which could be forthcoming using upcoming galaxy surveys such as DESI with Planck or the potential sky survey PIXIE.  If this \cii\ measurement is confirmed, it will open up a new window into large-scale structure, even up through the epoch of reionization.

\section*{Acknowledgments}
We thank J.~Aguirre, A.~Lidz, and A.~Myers for helpful comments.  AP was supported by a McWilliams Fellowship of the Bruce and Astrid McWilliams Center for Cosmology.  SH is supported by DOE and NSF AST1412966.

Part of the research described in this paper was carried out at the Jet Propulsion Laboratory, California Institute of Technology, under a contract with the National Aeronautics and Space Administration.
 
This work is based on observations obtained with \emph{Planck} (\url{http://www.esa.int/Planck}), an ESA science mission with instruments and contributions directly funded by ESA Member States, NASA, and Canada.
 
Funding for SDSS-III has been provided by the Alfred P. Sloan Foundation, the Participating Institutions, the National Science Foundation, and the U.S. Department of Energy Office of Science. The SDSS-III web site is \url{http://www.sdss3.org/}.

SDSS-III is managed by the Astrophysical Research Consortium for the Participating Institutions of the SDSS-III Collaboration including the University of Arizona, the Brazilian Participation Group, Brookhaven National Laboratory, Carnegie Mellon University, University of Florida, the French Participation Group, the German Participation Group, Harvard University, the Instituto de Astrofisica de Canarias, the Michigan State/Notre Dame/JINA Participation Group, Johns Hopkins University, Lawrence Berkeley National Laboratory, Max Planck Institute for Astrophysics, Max Planck Institute for Extraterrestrial Physics, New Mexico State University, New York University, Ohio State University, Pennsylvania State University, University of Portsmouth, Princeton University, the Spanish Participation Group, University of Tokyo, University of Utah, Vanderbilt University, University of Virginia, University of Washington, and Yale University.

\bibliographystyle{mnras}
 \newcommand{\noop}[1]{}

\label{lastpage}
\end{document}